\documentclass[times]{GonnellaLab-StyleArXiv}

\usepackage{blindtext}
\usepackage[overload]{textcase}
\usepackage{amsmath,amssymb}
\usepackage{lmodern}
\usepackage{iftex}
\usepackage{enumitem}
\usepackage{pifont}
\usepackage{booktabs}
\usepackage{multicol}
\usepackage[T1]{fontenc}
\usepackage[utf8]{inputenc}
\usepackage{textcomp} 
\usepackage{color}
\usepackage{fancyvrb}
\usepackage{xcolor}
\usepackage{natbib}
\usepackage{multirow}

\IfFileExists{upquote.sty}{\usepackage{upquote}}{}
\IfFileExists{microtype.sty}{
  \usepackage[]{microtype}
  \UseMicrotypeSet[protrusion]{basicmath} 
}{}
\makeatletter
\@ifundefined{KOMAClassName}{
  \IfFileExists{parskip.sty}{%
    \usepackage{parskip}
  }{
    \setlength{\parindent}{0pt}
    \setlength{\parskip}{6pt plus 2pt minus 1pt}}
}{
  \KOMAoptions{parskip=half}}
\makeatother
\IfFileExists{xurl.sty}{\usepackage{xurl}}{} 
\IfFileExists{bookmark.sty}{\usepackage{bookmark}}{}
\hypersetup{
  hidelinks,
  pdfcreator={LaTeX via pandoc}}
\DefineVerbatimEnvironment{Highlighting}{Verbatim}{commandchars=\\\{\}}

\ifLuaTeX
  \usepackage{selnolig}  
\fi

\begin{document}

\leadauthor{Gonnella}

\title{EGC: a format for expressing prokaryotic genomes content expectations}

\shorttitle{Expected Genome Contents format (EGC)}

\author[1,2\space \Letter]{Giorgio Gonnella}

\affil[1]{Center for Bioinformatics (ZBH), Universität Hamburg, Bundesstrasse 43, 20146 Hamburg}
\affil[2]{Institute for Microbiology and Genetics, Georg-August-Universität Göttingen, Goldschmidtstr. 1, 37077 Göttingen}

\maketitle

\begin{abstract}
The number of available genomes of prokaryotic organisms is rapidly growing enabling comparative genomics studies. The comparison of genomes of
organisms with a common phenotype, habitat or phylogeny often shows that these genomes share some common contents. 

Collecting rules expressing common genome traits depending on given factors is useful, as such rules could be used for quality control or for identifying interesting exceptions and formulating hypothesis.
Automatizing the rules verification using computation tools requires the definition of a representation schema.

In this study, we present EGC (Expected Genome Contents), a flat-text file format for the representation of expectation rules about the content of prokaryotic genomes. A parser for the EGC format has been implemented using the TextFormats software library, accompanied by a set of related Python packages.
\end {abstract}


\begin{corrauthor}
    giorgio.gonnella\at uni-goettingen.de
\end{corrauthor}

\begin{multicols}{2}
The application of comparative genomics techniques to the growing amount of
available prokaryotic genomes create expectations about the
genetic contents of organisms sharing a particular trait,
living in a common environment, or descending from a common ancestor.
Whenever expectations arise, it is interesting to verify if these hold
whenever new data becomes available. Thereby, unexpected results can
sometimes be an indication of low quality, or in other cases, more
interestingly, be confirmed and become the primer of new scientific
theories.

In order to automatize the collection and verification of rules of
expectations, it is necessary to create a suitable representation system.
In a recent study \citep{unambiguous}, we introduced a framework for
expressing such rules
consisting in the analysis of the logical structure of an expectation into its structural components
(definition of groups of organisms, definition of genome contents,
structure of different type of rules) and the introduction of
ontology systems to be used in the different definitions.
The system represents a foundation for representing rules, but does
not include concrete representation conventions.

In the present study we introduce a representation system for expectations
about the contents of prokaryotic genomes, build on that foundation,
in the form of a flat text format, called EGC (Expected Genome Contents).
The format has been implemented as a specification for the TextFormats
library \citep{textformats}. The implementation is complemented by several
Python packages. TabRec is a tools collection for handling files in
tabular records formats, including EGC. Conventions were developed
for representing particular types of data, accompanied by Python packages
for handling such data: Lexpr, for the logical expressions in the
definition of combined organism groups, Fardes, for the description of
genome feature arrangements, and TabRecPath, an addressing system for tabular
record formats, used in EGC for specifying the usage contexts of
user-defined tags and external links.

\begin{table*}
\centering
\begin{tabular}{lllp{10cm}}
\toprule
\textbf{Scope} & \textbf{Record type} &  & \textbf{Description}\\
\midrule
& \\[-3mm]
\multirow{2}{*}{\textit{Organism Groups}} &
\multirow{2}{*}{Group} &
\multirow{2}{*}{G} & groups of organisms
(e.g. by taxonomy, habitat, phenotype) for
which the rules are defined, or which are compared in rules \\
\midrule
\multirow{3}{*}{\textit{Genome Contents}}
& Genome Content Unit (GCU) & U & element of the genome sequence and/or annotation \\
& Attribute & A & measurable entity based on one or multiple
  genome content units \\
& Model & M & reference to an external model useful for the identification of
  GCUs \\
\midrule
\multirow{2}{*}{\textit{Expectation Rules}}
& Value rule & V & expectation of the value of an attribute based on reference values \\
& Comparative rule & C & expectation of the relation of the values of an
attribute in two groups \\
\midrule
\multirow{3}{*}{\textit{Sources}}
& Document & D & reference to an external text document, from which rules
are derived \\
& Text Snippet & S & snippet of a document text from which rules are derived \\
& Table & T & name of a table in a document from which rules are derived \\
\midrule
\multirow{2}{*}{\textit{Metadata}}
& External resource & X & database or ontology, of which elements are used
in other records \\
& Tag specification & Y & usage semantics and format of user-defined tags \\[1mm]
\bottomrule
\end{tabular}
\caption{List of record types in EGC in five scopes.}
\label{tab:RecordTypes}
\vspace{5mm}
\end{table*}

\section{Design principles of the EGC file format}

The EGC format was designed based on the same structure
of the GFA2 format \citep{gfa2}.
I.e. it consists of a list of records of different types,
where each record is expressed in a single line of the format. The fields
of each record are tab-separated.

The type of record is expressed in the first field and consists of a single
upcase letter. The following fields are in fixed number and format, different
for each type of record. If the content of a field is optional, a single dot
(\texttt{.}) is used to indicate the absence of information (placeholder).
After the fixed fields (positional fields), the line can contain any number of
additional fields, called tags. These have a semantic and datatype which is
defined using the system originally used in the SAM format: a two letter tag
name, a single letter tag datatype symbol, and the tag content are
separated by colons (\texttt{:}) and can include any spacing characters
except newlines and tabs.

The format described until this point reflects the same conventions used
in SAM and GFA records. Additionally, in EGC, each record may contain a last
field starting with a \texttt{\#}, which is handled as a record comment.

A specification of the format in the TextFormats Specification Language has been implemented, based on the design principles illustrated in \citet{tfslnewspec}.

Information which applies to different kind of records include identifiers and names, links to external resources (several records can be linked to external database, ontologies, scientific literature or web resources), tags (used to add further optional information to records) and comments (free texts added to different kind of records).

In some kind of records, both an identifier and a name or a description are present.
The main difference between these fields is their purpose. Identifiers are compact strings
whose main feature is stability, and are used as keys, i.e. allow a record to be referred from others. Conversely, names or descriptions are supposed to be more verbose texts, which may be
subject to change without impacting their connections from other records.

For sake of clarity and to allow combining multiple identifiers,
they only consists of letters,
numbers and underscores. Instead, names and descriptions can also include spacing and other
symbols (except tabs and newlines).

Each line in EGC can contain a free text comment, given after the last positional
field or tag (if any tag is present). The comment field is preceded by a field
separator (tab) and starts with a prefix consisting of \texttt{\#} and a space.
Comments may not contain tabs and end at the end of the line.

\section{Overview of the record types}

The format includes different type of records, summarized in Table \ref{tab:RecordTypes}. The different record types include the definition
of groups (G records), of genome contents (A, U and M records), of the
expectation rules (V and C records), as well as the documentation
of the textual sources from which the rules are derived (D, S and T records)
and optional metadata (X and Y records).

\section{Organism groups definition}

The definition of organism groups follows the conventions described in
\citet{unambiguous}. Records for the definition of groups have
the record type \texttt{G}. The record contain, in the order, the record type (G), a group identifier, a group name, the group type and a definition.
Examples are given in Figure \ref{Fig:G_examples}.

\begin{figure*}
\begin{Verbatim}[frame=single]
G  G_bacteria        Bacteria        taxonomic  taxid:2       TR:Z:domain
G  G_Xanthomonas     Xanthomonas     taxonomic  taxid:338     TR:Z:genus
G  G_Rickettsia      Rickettsia      taxonomic  taxid:780     TR:Z:genus
G  G_Thaumarchaeota  Thaumarchaeota  taxonomic  taxid:651137  TR:Z:phylum
G  G_copiotrophic    copiotrophic    nutrients_level_requirement \
                                       Wikipedia:Copiotroph   XR:O:ENVO:00002224
G  G_oligotrophic    oligotrophic    nutrients_level_requirement  \
                                       Wikipedia:Oligotroph   XR:O:ENVO:00002223
G  G_copiotrophic_B  copiotr. bact.  combined   G_copiotrophic & G_bacteria
G  G_oligotrophic_B  oligotr. bact.  combined   G_oligotrophic & G_bacteria
\end{Verbatim}
\caption{Examples of definitions of organisms groups in EGC format,
using G records. For ease of representation, some lines have been wrapped.}
\label{Fig:G_examples}
\end{figure*}

The group identifier is used in cross-references to groups from other records, is unique among all defined groups, and consists of letters, numbers and underscores only. The group name is a more
descriptive text, e.g.\ the scientific name of a taxon.

Groups are classified in \textit{group types}, whose definition is
given in the Prokaryotic Group Types Ontology \citep{unambiguous} (group types must be leaf nodes of the subtree under the \texttt{group\_types\_category} term).

The \textit{group definition} aims at providing a way to determine an
exact set of organisms contained in the group. This can be done
as a reference to an external database or ontology. For example in the case of taxonomic groups, the definition contains an ID of the NCBI taxonomy database (in the form \texttt{taxid:NNNNN}). When no external source is available, the definition may be provided as the string \texttt{def:} followed by a free text description.

For \textit{derived groups} (combinations or inversion of groups) the
identifiers of other groups are joined by the
logical operators \texttt{\&} (and), \texttt{|} (or), \texttt{\!} (not), possibly using round parentheses for indicating precedence.
Circular definitions are forbidden.

Tags can be added to the group records. The predefined tag \textit{XR} can be used to provide links to external resources, related to the group definition and is recommended when using \texttt{def:}.

\begin{figure*}
\begin{Verbatim}[frame=single]
U  Uo_CypX  ctg:ortholog_group:COG  COG2124    CypX      Cytochrome P450
U  Uo_aCOG  ctg:+ortholog_group:COG *          .         assigned to any COG
A  Ac_CypX  Uo_CypX                 count_relative:Uo_aCOG

U  Uy_T3SS  set:gene_system         .          T3SS      Type III secretion system
A  Ap_T3SS  Uy_T3SS                 complete_presence

U  Ug_16S   specific_gene           .          16S_rRNA  16S rRNA gene
U  Ug_23S   specific_gene           .          23S_rRNA  23S rRNA gene
U  Ug_5S    specific_gene           .          5S_rRNA   5S rRNA gene
U  Ur_rbs   set!:arrangement        Ug_16S,<>,Ug_23S,><,Ug_5S    .    \
                            16S rRNA separated from 23s rRNA gene and 5S rRNA gene
A  Ap_rbs   Ur_rbs                 complete_presence

U  Up_hxr   specific_protein        .          .         hydroxylamine oxidoreduct.
M  Up_hxr   InterPro                IPR010048  Hydroxylam_reduct
U  Uh_hxr   ctg:protein_homologs    homolog:Up_hxr    .    \
                                               hydroxylamine oxidoreductase homolog
A  Aph_hxr  Uh_hxr                  presence!replicon_type:plasmid
\end{Verbatim}
\caption{Examples of definitions of genome contents in EGC format. For ease of representation, some contents have been abbreviated and some lines wrapped.}
\label{Fig:GCU_examples}
\end{figure*}

\section{Genome contents definition}
The framework for the definition of genome contents is
described in \citet{unambiguous} and makes
use of definitions given in the Prokaryotic Genome Contents Definition Ontology (PGTO; \citet{unambiguous}). The genome attributes are measurable quantities defined in records of type \texttt{A}
as the value (presence/absence, relative or
absolute count) of a measurement (observation, computation, prediction) of entities belonging or derived from the
genome sequence or annotation (termed genome content units, GCUs,
defined in records of type \texttt{U}), in the entire
genome or a region thereof. Finally, feature model records (M)
are a means to provide further information about GCUs, for
identifying them in a genome.
Examples of genome contents definitions, including U, M and A records,
are given in Figure \ref{Fig:GCU_examples}.

\subsection{Genomic content units}

Genome content units (GCUs) records contain, in the order: the record type (\texttt{U}), GCU identifier (an identifier, unique among all U records), unit type, definition, symbol and description.

\subsubsection{Unit type}

Regarding the unit type, it shall be distinguished between simple units, category units and sets units.
In the case of simple units (i.e. single, specific features, which are not considered as composed of multiple parts),
the type consists of a single identifier.

Category units are groups of features which share some aspects, e.g. derivation from a common ancestor (e.g. group
of ortholog genes), structure (e.g feature types) or function. They can be defined by enumerating all its members
(in which case, the type identifier is prefixed by \texttt{ctg!:}), or by other means (in which case the prefix
is \texttt{ctg:}), such as referring to an external
database accession, by describing the category, or by enumerating its members.
In case the components are enumerated, the prefix will be \texttt{ctg!:}.
The identifier itself can either refer to the category in its entirety (e.g. \texttt{ctg:function}), or rather
identify each element of it, in which case it is prefixed by \texttt{+} (e.g. \texttt{ctg:+specific\_protein}).
When an external database is used for the definition, a resource identifier is added as suffix after a further (\texttt{:}).
An example is \texttt{ctg:ortholog\_group:COG} for ortholog groups defined in the COG database \citet{Galperin2020}).

Sets of units are units, in which distinct sub-units or components can be distinguished (e.g. their presence or absence is evaluated
separately). The same notation conventions
are used, as for categories, with the prefix \texttt{set!:} if the members are enumerated, \texttt{set:} otherwise,
and preceding the identifier by a \texttt{+} (e.g. \texttt{set!:+specific\_protein}), if this refers to type of the single elements.
Also in this cases, entries of external resources can be provided as definitions, and the resource identifier is added as a suffix.

Different types, following these conventions, and their use cases are summarized in Table \ref{tab:UnitTypes}.
Specific sequence letters or genes and proteins are defined by their name. Categories are defined by function
or homology to other units or membership in groups, defined by homology, such as
gene ortholog groups, e.g.\ COG \citep{Galperin2020}) or protein families, e.g.\ PFAM \citep{Mistry2020}.
Set are defined e.g. by their common transcriptional regulation (operons),
by their proximity (gene clusters, gene islands), or common functions
(gene systems, metabolic pathways).

\subsubsection{Definition, symbol and description}

Three fields are available, to describe the contents of a unit:
thereby ``definition'' has a format appropriate to the type, as described below,
``symbol'' is a compact identifier, by which the unit is known in literature or a databases,
and ``description'' is a free-text description.
At least one of these must be provided (the remaining are filled with a single dot).

For enumeration types (with prefix \texttt{ctg!:} or \texttt{set!:}), the definition
field contains a comma-separated list of IDs of other units defined in the data set.
An exception are \texttt{set!:arrangements}
for which the more complex \textit{fardes} format has been developed, which is described
in the next section (in the simplest case, it is also a comma-separated list of unit IDs).

For non-enumeration types, in which a resource identifier is provided (e.g. \texttt{ctg:ortholog\_group:arCOG},
the definition field contains the accession or term ID in that resource. For \texttt{+} types
(e.g. \texttt{ctg:+ortholog\_group}:COG), a comma-separated list of accessions or term IDs is given,
or the special symbol \texttt{*}, meaning ``any members of the resource''.

In other cases, when a unit definition is given by providing a link to an external document
(e.g. a scientific article), the
definition field contains \texttt{ref:} followed
by the external resource link (e.g. \texttt{ref:doi:...}).

Specific genes and proteins are defined by their names or symbols, which can be often be ambiguous
(e.g.\ in different organisms, the same symbol may refer to different genes, or
the same gene may have different names). To provide additional information,
model records can be added, which refer to external databases and give possibility
to the user to operationally transform the specific unit definition into clearer category
definitions.  These records have record type \texttt{M}, followed by the unit ID,
the ID of the external database ID, the model ID and name
its name in the database.

\subsection{Features arrangement description Format (Fardes)}

For units describing feature arrangements, the definition field 
content is encoded using a string notation here introduced,
named \textit{Fardes} (feature arrangement description).

A fardes string consists of a list of
named feature and, optionally, interval specifiers.
Named features are identified by an ID, which in EGC must be the ID
of a GCU definition record (type U). If they are optional,
their ID is followed by a question mark (e.g. \texttt{U1?}).

Interval specifiers are optional and describe what is in between two
subsequent named features: sequence length, number and type of features.
A full specifier has the form \texttt{a:b(type)[c:d]} and can be shortened
if defaults apply. Thereby, 
\texttt{a:b} is the range of number of features in the interval, \texttt{type} their
feature types, \texttt{c:d} the length of the interval sequence.
Instead of the \texttt{a:b} and \texttt{c:d} form, a number preceded by a comparison
operator (e.g. \texttt{>2}) can be given. The number of features can be omitted
if it is just \texttt{>=0}.
For the length a \texttt{~} symbol can be used
to indicate an approximate length, and units can be optionally used
(e.g. \texttt{~3Mb}). The length can be omitted if it is \texttt{[>=0]}.
If a single number is given instead of a range, it is used
as both minimum and maximum of the range.
If no interval specifier is used, the default is \texttt{0:0[0:*]}, i.e. there
no features between the two named features and no interval length constraint.

Special interval specifiers are used to indicate fuzzy terms, such as
overlapping (\texttt{\&}),
near (\texttt{><}), distant (\texttt{<>}), distant but in the same molecule
(\texttt{<.>}) or in different molecules (\texttt{<|>}).

The relative strand arrangement can be specified, by prefixing the ID of
a feature with \texttt{>}. This then becomes the reference point, and subsequent
features ID prefixed with \texttt{=} and \texttt{\^} indicate the same or the opposite strand to it. The first feature of the arrangement and the first feature
after a \texttt{<|>} specifier have an implicit \texttt{>}.

An example of arrangements string in fardes format is
\texttt{U1,U2?,>1(tRNA;rRNA),U3,1:3,=U4,<>,}
\texttt{>U5,[~1kb],\^U6,<|>,U7}.
Its meaning is: U1 is optionally followed by U2, then at least one tRNA
or rRNA gene, but possibly more, then U3, one to three other features
and U4, which must be on the same strand as U1. Distant from this cluster,
is U5, and about 1 kb downstream U6, which is on the opposite strand of U5.
Finally U7 is on another molecule from the previously named features.

\subsection{Genome attributes}

The genome attributes are the entities that can be measured in a genome,
whose values are compared in rules to reference values or to other genomes.
They always refer to a genome content unit.

The records for the definition of an attribute contains the record type (A),
an identifier (unique among all A records), the genome content unit ID,
and an attribute definition field.

The contents of the definition fields specify the measurement mode, the
reference unit for relative measurements and the genomic region where
the measurement is computed.
Table \ref{tab:AttributeModes} summarizes the possible choices.
For specific units and categories, these include the presence, count and total sequence length
of specific units and categories. Count and length can be relative, in which
case a reference unit ID must be provided.
For sets, it must always be specified if the mode is intended to be applied
to the complete set (prefix \texttt{complete\_}) or to any of the members
(prefix \texttt{members\_}).

By default, the value is computed from the entire genome, but a region
can be specified, in terms of molecule type or name, using the syntax
summarized in Table \ref{tab:AttributeRegion}.

\begin{table*}
\centering
\begin{tabular}{llp{9cm}}
\toprule
\textbf{Kind} & \textbf{Type} & \textbf{Description} \\
\midrule
& \\[-3mm]
Sequence &\\
& \\[-3mm]
\ \ Specific
  & \texttt{base} & single base of DNA/RNA \\
  & \texttt{amino\_acid} & an amino-acid \\
& \\[-3mm]
\ \ Category
  & \texttt{ctg!:+base} & one of a set of bases \\
& \\[-3mm]
\midrule
& \\[-3mm]
Annotation &\\
& \\[-3mm]
\ \ Specific
  & \texttt{specific\_gene} & gene with given name \\
  & \texttt{specific\_protein} & protein with given name \\
& \\[-3mm]
\ \ Category
  & \texttt{ctg:function} & function of a gene/protein,  e.g.\ enzymatic activity \\
  & \texttt{ctg:feature\_type} & any feature of a given type, e.g. CRISPR \\
  & \texttt{ctg:gene\_homologs} & any homolog of a unit of type \texttt{specific\_gene} \\
  & \texttt{ctg:ortholog\_group} & group of ortholog genes \\
  & \texttt{ctg:ortholog\_group:}\textit{<ID>} & ortholog group from database \textit{<ID>} \\
  & \texttt{ctg:ortholog\_groups\_category} & multiple ortholog groups with similar functions \\
  & \texttt{ctg:protein\_homologs} & any homolog of a unit of type \texttt{specific\_protein} \\
  & \texttt{ctg:family\_or\_domain} & protein family or group of proteins contanining a given domain\\
& \\[-3mm]
\ \ Set
 & \texttt{set:metabolic\_pathway} & set of gene/protein functions,
 which together create a pathway \\
 & \texttt{set:protein\_complex} & protein whose units are coded by multiple genes \\
 & \texttt{set:+specific\_gene} & generic set of multiple genes \\
 & \texttt{set:gene\_system} & genes with a common function (can be distant to each other)\\
 & \texttt{set:gene\_cluster} & genes with a common function and next to each other \\
 & \texttt{set:operon} & genes in an operon \\
 & \texttt{set:genomic\_island} & genes in a genomic island \\
 & \texttt{set:arrangement} & relative positioning of genes \\[2mm]
\bottomrule
\end{tabular}
\caption{Overview of the types of genome content units in EGC.
The list is not exhaustive and further types can be defined if necessary.
The set and category definitions can be provided as enumeration by using the prefix
\texttt{set!}/\texttt{ctg!} instead of \texttt{set}/\texttt{ctg}.}
\label{tab:UnitTypes}
\vspace{5mm}
\end{table*}
 
\begin{table*}
\centering
\begin{tabular}{lllp{10cm}}
\toprule
\textbf{Kind} & \textbf{Type} & \textbf{Datatype} & \textbf{Description} \\
\midrule
& \\[-3mm]
Specific /
     &           \texttt{presence} & boolean &
                 presence of the specific feature, or of a member of the group \\
\ Category
     &           \texttt{count} & integer &
                 number of instances of the specific feature or category \\
     &           \texttt{length} & integer &
                 total sequence length of instances of the specific feature or category \\
     &           \texttt{count\_relative}:\textit{<ID>} & float &
                 \texttt{count}, divided by the number of instances of reference
                 unit \textit{<ID>} \\
     &           \texttt{length\_relative}:\textit{<ID>} & float &
                 \texttt{length} divided by the total length of instances of reference
                 unit \textit{<ID>} \\
& \\[-3mm]
\midrule
& \\[-3mm]
Set 
    & \texttt{complete\_...}      &         & any mode considering all members of the set, e.g.: \\
    & \ \ \texttt{complete\_presence} & boolean & presence of all members of the set \\
    & \ \ \texttt{complete\_count}    & integer & number of complete copies of the set \\
    & \texttt{members\_...}      &         & any mode considering any member of the set, e.g.: \\
    & \ \ \texttt{members\_presence}  & boolean & presence of at least one member of the set \\
    & \ \ \texttt{members\_count}     & integer & count of the instances of any member of the set \\
\bottomrule
\end{tabular}
\caption{Overview of the attribute definition field syntax for different kind
of units and measurement modes.}
\label{tab:AttributeModes}
\vspace{5mm}
\end{table*}

\section{Rules of expectations}

The purpose of the EGC format is to describe expectations about the genome contents
in given groups. The expectations can be given as relative to reference values,
described in records of type V (value), or relative to other groups,
described in records of type C (comparison). Examples of this records
are given in Figure \ref{Fig:rules_examples}.

The value records contain the record type (V), an identifier (unique among all
V and C records), a source, an attribute ID, a group descriptor, an operator
and a reference.

The comparison records contain the record type (C), an identifier (unique
among all V and C records), a source, an attribute ID, a first group descriptor,
an operator and a second group descriptor.

For comparing between two regions of the same genome (e.g. two chromosomes),
two distinct attributes are defined and their IDs are concatenated by a comma
in the attribute ID field. In this case both group descriptor will be identical.

The source consists of a single identifier of S or T record,
or a list, comma-separated.

\begin{figure*}[!b]
\begin{Verbatim}[frame=single]
C  C1  S1  A_rel_COG2124                G_copiotrophic_B    <<  G_oligotrophic_B
V  V1  S2  A_has_T3SS_sys               G_Xanthomonas:most  ==  True
V  V2  S3  A_has_rRNA_arr1              G_Rickettsia        ==  True
V  V3  S4  A_has_hydroxylam_reduct_hom  G_Thaumarchaeota    ==  False
\end{Verbatim}
\caption{Examples of definitions of expectation rules in EGC format}
\label{Fig:rules_examples}
\end{figure*}

The group descriptor is either just a group ID, indicating that the expectation
concerns all members of the group, or a group ID followed by 
a semicolon and one of the following terms: \texttt{rare}, \texttt{some},
\texttt{many} or \texttt{most}, or a symbol \texttt{>} or \texttt{<} and a percentage (e.g.\ \texttt{>90\%}).

Operators for exact comparisons and fuzzy comparison are indicated in
Table \ref{tab:RuleOperators}. 
The definition of exact values for the verification of rules including fuzzy operators
are left to the verification tools implementation.

\begin{table*}
\centering
\begin{tabular}{llp{10cm}}
\toprule
\textbf{Region} & \textbf{Type suffix} & \textbf{Description} \\
\toprule
& \\[-3mm]
Whole genome & \textit{none} & attribute measured in the whole genome\\
Molecule type & \texttt{!replicon\_type:}\textit{ID} &
 attribute measured only in replicons of the given type
(e.g. \texttt{chromosome} or \texttt{plasmid})\\
Specific molecule & \texttt{!}\textit{type}\texttt{:}\textit{name} &
 attribute measured in replicons of the given type
(e.g. \texttt{chromosome} or \texttt{plasmid})
and with the given name\\
\bottomrule
\end{tabular}
\caption{Overview of the attribute definition field suffix syntax for specifying
different regions of the genome.}
\label{tab:AttributeRegion}
\vspace{5mm}
\end{table*}

\begin{table*}
\centering
\begin{tabular}{llllp{8cm}}
\toprule
\textbf{Category} & \textbf{Values} & \textbf{Operators} & \textbf{N.\ ref.} & \textbf{Description} \\
\toprule
& \\[-3mm]
Exact & Numerical & \texttt{==}, \texttt{!=} & 1 & identity or difference from
given value \\
& & \texttt{<}, \texttt{>}, \texttt{>=}, \texttt{<=} & 1 & numerical comparison with given value\\
& Numerical & \texttt{in\_range} & 2 & 
in the range for which the (inclusive) minimum and maximum limits of a reference range are given\\
& Boolean & \texttt{==} & 1 & compare a boolean value to True or False. \\
\midrule
Fuzzy & Numerical & \texttt{>}\texttt{>}, \texttt{<}\texttt{<} & 1 &
much larger or smaller than the reference\\
& & \texttt{>\~}, \texttt{<\~} & 1 &
 slightly larger or smaller than the reference\\
& Numerical & \texttt{level} & 1 & rough level category,
i.e. \texttt{none\_or\_low} \texttt{low} or \texttt{high} \\
\bottomrule
\end{tabular}
\caption{Operators which can be used in value rules.}
\label{tab:RuleOperators}
\vspace{5mm}
\end{table*}

\section{Documentation of textual sources}

The EGC format aims at documenting the sources of the expectation rules.
The system includes records for storing references to documents (D) and snippets of text (S) or references to tables (T) inside the documents.
An example is given in Figure \ref{fig:srcexamples}.

\begin{figure*}
\begin{Verbatim}[frame=single]
D  pmid:19805210  https://www.pnas.org/doi/abs/10.1073/pnas.0903507106
D  pmid:32983016  https://www.frontiersin.org/articles/10.3389/fmicb.2020.01991
D  pmid:15317790  https://journals.asm.org/doi/10.1128/JB.186.17.5842-5855.2004
D  pmid:25587132  https://www.pnas.org/content/112/4/1173.long
S  S1  pmid:19805210  In addition, cytochrome P450 genes (COG2124) are present in
                      S. alaskensis RB2256 (six copies) and absent in P. angustum
                      S14, and the high frequency by which these genes occur is a
                      conserved feature of oligotroph genomes.
S  S2  pmid:32983016  The assessment of 133 pathogenicity-related genes identified
                      that the three Xanthomonas strains (GW, SS and SI) was devoid
                      of the T3SS that is critical for pathogenicity of most
                      Xanthomonas species.
S  S3  pmid:15317790  As in the other rickettsiae, the 16S rRNA gene was separated
                      from the 23S and 5S rRNA genes.
S  S4  pmid:25587132  As with all previously sequenced Thaumarchaeota, no
                      hydroxylamine oxidoreductase homologs were identified.
\end{Verbatim}
\caption{Examples of documentation of sources in EGC format.
Note that the S records texts have been displayed wrapped to multiple lines
in order to display it here, but is in reality all contained in a single line.
The sentences are extracted from \citet{Lauro2009} (S1), \citet{Li2020} (S2),
\citet{McLeod2004} (S3) and \citet{Santoro2015} (S4).}
\label{fig:srcexamples}
\end{figure*}

The document description records contain the following fields, in the order: record type (D), document ID, full-text link.
Thereby the document ID is a link to an external resource listing the document. For example, the Pubmed ID can be used, by preceding it with the prefix \texttt{pmid:}, or the DOI
by preceding it with the prefix \texttt{doi:}.
The full-text link is an optional field.
It can be set to a placeholder character \texttt{.} if not necessary.
(since e.g. it is trivial to find a document if the DOI is provided).
Each record of type \texttt{S} or \texttt{T} refers to a record of type \texttt{D} (many can refer to the same one).

The text snippets records contain, in the order: record type (S),
snippet ID, document ID, text snippet. The snippet ID is an identifier, unique among all S records. The document ID refers to a D record.
The text snippet is the text of the sentence(s) or part thereof, from which a rule can be extracted, without any newline or tab character.

The structure of the table records is similar, but the contents of the table are not included, replaced instead by a reference to it. They contain: record type (T), table ID (a unique identifier among all T), document ID (referring to a D record),
table reference (e.g. table number in the document).

\section{Tags}

EGC records support the use of tags, which provide a flexible way to store additional information. Tags, first introduced by the SAM format \citep{sam}
and later extended to other formats, such as VCF \citep{vcf} and GFA
\citep{gfa2}, are optional fields to store
information about the record that is not captured in the other fields.

Tags in the EGC format use the same formatting as SAM tags:
each tag consists of a name of two letters, representing the
semantics of the information, followed by a type code, and a value of that
type.

In EGC only uppercase tag names are allowed and the type codes described in
Table \ref{tab:TagTypeCodes} are supported. Compared to SAM, the JSON code \texttt{J} has
been introduced (which also exists in GFA2), as well as specialized codes for
lists of strings (\texttt{L}) and of ontology terms (\texttt{O}).

\begin{table*}
\centering
\begin{tabular}{llp{12cm}}
\toprule
\textbf{Type code} & \textbf{Purpose} & \textbf{Format}\\
\midrule
& \\[-3mm]
\texttt{Z} & generic string & not containing tabs and newlines \\
\texttt{A} & character & single character, except tab and newline \\
\texttt{J} & JSON & not containing tabs and newlines \\
\texttt{i} & integer value & signed or unsigned integer \\
\texttt{f} & floating-point value & IEEE 754 text representation, single precision\\
\texttt{L} & list of strings & semicolon-separated, elements may not contain
semicolons, tabs, and newlines \\
\texttt{O} & list of ontology terms & semicolon-separated, no newlines, tabs, \texttt{;}, \texttt{:}, \texttt{\#}\\
& & 
format: \texttt{ont\_pfx:term\_id\#term\_label} or \texttt{ont\_pfx:term\_id};\\
\texttt{H} & byte array & in hex format, as in SAM format specification (*) \\
\texttt{B} & numeric array & as in SAM format specification (*) \\
\bottomrule
\end{tabular}
\caption{Tag-type codes in the EGC format. (*) SAM format specification, see reference: \citet{samspec}}
\label{tab:TagTypeCodes}
\vspace{5mm}
\end{table*}

The list of predefined tags is reported in Table \ref{tab:PredefinedTags}.
User-defined tags can be used. In this case a tag definition records (type \texttt{Y}) can be employed to document the semantics and format of such tags. These
records contain, after the record type, the tag name (a 2 letter code),
the tag type, usage contexts (as explained later), semantics and format.
An example is given in Figure \ref{fig:meta_examples}.

\begin{table*}
\centering
\begin{tabular}{llllp{5cm}p{4cm}}
\toprule
\textbf{Name} & \textbf{Label} & \textbf{Type}
& \textbf{Context} & \textbf{Description} & \textbf{Value format} \\
\midrule
& \\[-3mm]
XR & eXternal Resource & Z & G & link related information & external resource format \\
XL & eXternal List & Z & G & link members enumeration & external resource format \\
XD & eXternal Definition & Z & G:combined & link definition & external resource format \\
TS & Taxon Species & Z & G:strain & link species of a strain & taxid:{} \\
TG & Taxon Genus & Z & G:strain & link genus of a strain & taxid:{} \\
TR & Taxonomic Rank & Z & G:taxonomic & rank of the taxon & rank from NCBI taxonomy \\
\bottomrule
\end{tabular}
\caption{List of predefined tags in the EGC format. These tags may be used directly, while for using any other tag, a tag definition record must be included in the data.}
\label{tab:PredefinedTags}
\vspace{5mm}
\end{table*}

\begin{figure*}
\begin{Verbatim}[frame=single]
X  Wikipedia  English-language Wikipedia  G.definition;U.definition \
                https://en.wikipedia.org/wiki/\{\}  https://en.wikipedia.org  .
Y XR  Z  external resource with related information  external resource link
\end{Verbatim}
\caption{Examples of metadata records for describing external resources (X) and
tags (Y). The X record content has been wrapped in order to allow
displaying its whole content, but is in fact contained in a single line.}
\label{fig:meta_examples}
\end{figure*}

\section{References to external resources}

Some fields contains references to external resources, such as 
an ontology term, or an item in an external database.
These are given in the form:\\
\texttt{resource\_prefix:item[\#location][!term]}\\
Thereby:

\begin{itemize}
\item the resource prefix identifies a resource (website, dictionary, ontology,
  database or similar)
\item the item identifies the relevant element of the resource, e.g. term in
  a dictionary or ontology, or record in a database
\item the following parts are optional
\item the \texttt{\#} part is used for providing further information about the
  location of the item inside the pointed document, if needed; for URLs
  this is part of the URL itself.
\item the \texttt{!} part can be used for resources such as ontologies and dictionaries
  for annotating the name of the linked term; this is especially useful when
  the item ID is unrelated to the term itself, such as it is usually the case
  in ontologies.
\end{itemize}

Some external resources are predefined, i.e.\ have the
same name, point
to the same URL and share the same allowed usage context in all documents.
The list of predefined external resources is reported in Table \ref{tab:PredefinedExternalResources}.
This includes some common databases and all ontologies
registered in the OBO foundry (and listed at https://ontobee.org/).

\begin{table*}
\centering
\begin{tabular}{lll}
\toprule
\textbf{Prefix} & \textbf{Name} & \textbf{Item URL [Homepage URL]}\\
\midrule
& \\[-3mm]
\texttt{taxid} & NCBI taxonomy database & https://www.ncbi.nlm.nih.gov/taxonomy/?term=\{\} \\
& & [ \url{https://www.ncbi.nlm.nih.gov/taxonomy} ]\\
\texttt{http} & generic http URL & http:\{\} \\
\texttt{https} & generic https URL & https:\{\} \\
\texttt{ftp} & generic ftp URL & ftp:\{\} \\
\texttt{doi} & Digital Object Identifier & https://doi.org/\{\} \hspace{2.3cm}  [ \url{https://doi.org} ] \\
\texttt{bacdive} & Strain in BacDive database & https://bacdive.dsmz.de/strain/\{\} \ \  [ \url{https://bacdive.dsmz.de/} ] \\
\texttt{biosample} & NCBI BioSample database & https://www.ncbi.nlm.nih.gov/biosample/?term=\{\} \\
& & [ \url{https://www.ncbi.nlm.nih.gov/biosample/} ] \\
\texttt{dsmz} &  Strain in DSMZ database & https://www.dsmz.de/collection/catalogue/details/culture/\{\} \\
& & [ \url{https://www.dsmz.de/} ] \\
\texttt{sctid} & SNOMED Clinical Terms & https://browser.ihtsdotools.org/?perspective=full\&conceptId1=\{\} \\
& & [ \url{https://www.snomed.org/} ]\\
OBO pfx (*) & OBO foundry ontology & http://purl.obolibrary.org/obo/PFX\_\{\} \\
& & [ \url{https://ontobee.org/ontology/} ] \\
\bottomrule
\end{tabular}
\caption{List of predefined external resources in the EGC format. The item URLs are URL patters containing \{\} as a placeholder, which
is substituted in the URL with the ID of the element. 
(*) For OBO ontologies the prefix is the one registered
in the OBO foundry (https://obofoundry.org/), and is indicated
as {PFX} in the item URL pattern and homepage URL.}
\label{tab:PredefinedExternalResources}
\vspace{5mm}
\end{table*}

For other external resources, the user may want, for clarity, 
include a definition record. These records have type X and contain, in the order,
a resource prefix (unique among all resources), a resource name, the
usage contexts (as explained later), the URL pattern for addressing single items
(if applicable), the URL of the homepage of the resource, and a reference to a scientific article or other descriptive text. An example is given in Figure \ref{fig:meta_examples}.

\section{Usage contexts}

In the definition of tags and external resources, a usage context for the
defined items is specified. This is done by means of a fields and
records selection path format,
called \textit{TabRecPath} and presented here, which can be applied to any tabular records
files, i.e. files whose structure resembles that of EGC.

The mini-format can be used for addressing fields in a record,
e.g. necessary for specifying the allowed usage contexts
of external resources. It can also be used for addressing records,
e.g. necessary for specifying the usage contexts of tags.
Besides selecting a record type, record subtypes can also be selected,
if a subtype exists for a given record type.
In order to have a compact string representation,
multiple subtypes and field names or numbers can be specified.
Multiple addresses can be concatenated.

Records are addressed by a string in the form
\texttt{rt1;rt2:st1|st2;...} and fields are addressed by a string in the form
\texttt{rt1.f1;rt2:st1|st2.f3|f4;...}.
Thereby:
\begin{itemize}
\item \texttt{rt} are record types
  (e.g. \texttt{G} for groups)
\item \texttt{st} are the optional subtypes, can be added for records
  for which subtypes exists, e.g. EGC groups
  (e.g. \texttt{G:combined})
\item \texttt{f} (fields) are the numbers (1-based) or names of positional fields, or tag names (e.g. \texttt{G:strain.TS})
\item multiple addresses are separated by semicolons (\texttt{;})
\item multiple subtypes or field names are separated by vertical lines (\texttt{|})
\item identifiers of record types, subtypes and fields
  may only contain letters, numbers and underscores and may not
  start with a number
\end{itemize}

For example, the usage context:\\[1mm]
\texttt{G:strain.TS|TG;G:taxon|comb.XR}\\[1mm]
means that the corresponding feature can be used
in the fields \textit{TS} and \textit{TG} of the \textit{strain} subtype of \textit{G} records and in the \textit{XR} field of
the \textit{taxon} and \textit{comb} subtype of \textit{G} records.

\section*{Implementation}

The format has been implemented in Python packages. These include the
EGC specification package and additional packages for specific parts
of the format, which can be useful also in other contexts (Lexpr parser
for logical expressions used in the definition of combined groups,
Fardes for the features arrangement descriptions, TabRecPath for
selecting specific records and fields) and for handling files with
a similar structure to EGC (TabRec package).

\subsection{EGC TextFormats specification}

A parser for the format has been implemented as a specification for the TextFormats library. This allows to read and write the format from different programming languages, such as Nim, Python, C and C++. 
The specification is available in Github at
\url{https://github.com/ggonnella/egc-spec}.

The main module, which can be employed in external code using the TextFormats library is the \texttt{egc.tf.yaml}.

The different classes of records are defined in different separate
modules included in it, for the definitions of groups,
genome contents, textual sources,
expectation rules, tags, external resources links and
usage contexts.

\subsection{Lexpr parser}

The logical expressions contained in derived group definitions can
be parsed using the Lexpr package, implemented by creating a grammar
for the Python library Lark.

The package can be installed by \texttt{pip} and is available in Github
(\url{https://github.com/ggonnella/lexpr}).

\subsection{Fardes format}

The Feature Arrangements Description format is implemented using
Python and TextFormats in the Python package Fardes, installable by \texttt{pip}. The source code is available on Github (\url{https://github.com/ggonnella/fardes}).

The module \texttt{fardes/parser.py} in the package implements a parser
and validator for the format.

\subsection{TabRec CLI tools}

A Python package has been implemented for analysing
and editing files which has a structure similar to EGC (tab-separated
records). It can be installed by \texttt{pip} and the
source code is available in Github 
(\url{https://github.com/ggonnella/tabrec}).

Thereby, the \texttt{tabrec-analyse}
tool is a versatile CLI tool for the analysis of the
contents of a tabular records file, e.g. a EGC file. It allows to show
the values set and statistics about the
value distributions of single fields or all fields of records of
a given type in the file.

Other CLI tools in the package
allow extracting a few or all lines of a given record type,
or editing operations, such as swapping the contents of two fields,
limited to a given record type.

\subsection{TabRecPath format}

In the text, we described the
 TabRecPath format for the selection of records and fields in tabular
record formats, such as EGC.

A description of the format (\texttt{docs/TabRecPath.md}), 
a TextFormats specification file including examples
(\texttt{tabrec/data/tabrecpath.tf.yaml},
a Python implementation (\texttt{tabrec/path.py}),
and a CLI tool for extracting fields based on TabRecPath fields paths
(\texttt{bin/tabrec-extract})
are available in the TabRec github repository
(\url{https://github.com/ggonnella/tabrec}).

\section*{Discussion and Conclusion}

In this paper, we present a representation for rules of expectations
about the contents of prokaryotic genomes. In particular, a file format
was developed, named EGC (expected genome contents), on the base of a logical analysis of
such expectations, described in a separate manuscript \citep{unambiguous}.
The purpose of the format is to store rules of expectations, so that
a concrete verification of those rules becomes possible. 
To the best of our knowledge, this is the first representation available for this purpose.

The structure of the EGC format was based on existing bioinformatics
formats, such as GFA \citep{gfa1,gfa2}, which consists of tab-separated
files, where each line is a record, and where different record types
are possible, distinguished by the content of the first field.
Such structure has the advantage to be flexible and extensible and
more compact, compared to other solutions, such as JSON
\citep{pezoa2016foundations}.

We propose to use the term \textit{tabular records format} as a general term for this kind of formats and provide some command-line tools,
implemented in Python, which can be used for such formats
(TabRec), and an addressing system (TabRecPath) for referring to given fields or records (distinguished by their type and an optional subtype).
This system is inspired by similar systems for XML (XPath, \citet{xpath}) and JSON (JSONPath, \citet{jsonpath}).

Rules stored in EGC files
are likely to be derived
from scientific literature and the format aims at documenting
the sources exactly.
The structure of the format is flexible,
and new record types can be added to it, to extend is function to
the representation of rules derived differently.

Convention necessities arose, when developing the format. For example,
no notation was available, for expressing arrangements of
sequence features in a genome.
Thus, a notation for this purpose, named
Fardes (Features Arrangement Description),
was developed and implemented as
a separate Python package.

Furthermore, the definitions of organism groups and genomic contents
require references to elements of external resources. For this purpose
a link format similar to URIs \citep{BernersLee1998} was defined, unifying references to ontologies, databases, websites and dictionaries.
Additionally we provided a method for giving
optional descriptions of the resources themselves.

The formats adopts the tag system originally developed for SAM \citep{samspec}. The system was here modified by introducing new types
for lists of identifiers and ontology term links.
As for external resources, we provided a method for giving
optional description of the semantics and format
of user-defined tags.

To conclude, the format described in this article allow representing
a kind of data, expectations about genome contents in different
groups of organisms, which has
not been handled in other file formats before.
Thus, this study represents a practical and necessary
foundation for implementing
tools for the verification of such expectations.

\begin{acknowledgements}
Giorgio Gonnella has been supported by the DFG Grant GO 3192/1-1 ‘`Automated characterization of microbial genomes and metagenomes by collection and verification of association rules’’. The funders had no role in study design, data collection and analysis, decision to publish, or preparation of the manuscript.

The author would like to thank Serena Lam (University of G\"ottingen) for discussions regarding the expectation rules representation, which were helpful in the
format specification, and for providing the examples rules mentioned in the
text.
\end{acknowledgements}

\begin{contributions}
 These contributions follow the Contributor Roles Taxonomy guidelines: \href{https://casrai.org/credit/}{https://casrai.org/credit/}.
 Conceptualization: G.G.;
 Data curation: G.G.;
 Formal analysis:  G.G.;
 Funding acquisition:  G.G.;
 Investigation: G.G.;
 Methodology: G.G.;
 Project administration: G.G.;
 Resources: G.G.;
 Software: G.G.;
 Supervision: G.G.;
 Validation: G.G.;
 Visualization:  G.G.;
 Writing – original draft: G.G.;
 Writing – review \& editing: G.G.
\end{contributions}

\begin{interests}
 The authors declare no competing financial interests.
\end{interests}

\bibliography{references}

\end{multicols}

\end{document}